\documentclass[preprint]{aastex61}
\usepackage{verbatim} %%% 大段注销
\submitjournal{ApJ}
\shorttitle{Two-sided loop jets caused by magnetic reconnection between filamentary threads}
\shortauthors{Tian et al.}
\newcommand{\nfig}[1]{Figure~\ref{#1}}
\newcommand{\speed}[1]{#1 km~s${}^{-1}$}

\begin{document}

\title{Successive Two-sided loop Jets Caused by Magnetic Reconnection between Two adjacent Filamentary  Threads}
\correspondingauthor{Yuandeng Shen}
\email{ydshen@ynao.ac.cn}
\author{Zhanjun Tian }
\affiliation{Yunnan Observatories, Chinese Academy of Sciences,  Kunming, 650216, China }
\affiliation{State Key Laboratory of Space Weather, Chinese Academy of Sciences, Beijing 100190, China}\affiliation{University of Chinese Academy of Sciences, Beijing 100049, China}
\author{Yu Liu}
\affiliation{Yunnan Observatories, Chinese Academy of Sciences,  Kunming, 650216, China }
\affiliation{Center for Astronomical Mega-Science, Chinese Academy of Sciences, Beijing, 100012, China}
\affiliation{Key Laboratory of Geospace Environment, Chinese Academy of Sciences, University of Science \& Technology of China, Hefei 230026, China}
\author{Yuandeng Shen}
\affiliation{Yunnan Observatories, Chinese Academy of Sciences,  Kunming, 650216, China }
\affiliation{State Key Laboratory of Space Weather, Chinese Academy of Sciences, Beijing 100190, China}
\affiliation{Center for Astronomical Mega-Science, Chinese Academy of Sciences, Beijing, 100012, China}
\affiliation{Key Laboratory of Solar Activity, National Astronomical Observatories, Chinese Academy of Science, Beijing 100012, China}
\affiliation{Key Laboratory of Geospace Environment, Chinese Academy of Sciences, University of Science \& Technology of China, Hefei 230026, China}
\author{Abouazza Elmhamdi}
\affiliation{Department of Physics and Astronomy, King Saud University, P.O. Box 2455, 11451, Saudi Arabia}
\author{Jiangtao Su}
\affiliation{University of Chinese Academy of Sciences, Beijing 100049, China}
\affiliation{Center for Astronomical Mega-Science, Chinese Academy of Sciences, Beijing, 100012, China}
\affiliation{Key Laboratory of Solar Activity, National Astronomical Observatories, Chinese Academy of Science, Beijing 100012, China}
\author{Ying D. Liu}
\affiliation{State Key Laboratory of Space Weather, Chinese Academy of Sciences, Beijing 100190, China}
\affiliation{University of Chinese Academy of Sciences, Beijing 100049, China}
\author{Ayman. S. Kordi}
\affiliation{Department of Physics and Astronomy, King Saud University, P.O. Box 2455, 11451, Saudi Arabia}
%\email{ydshen@ynao.ac.cn} \email{tzhj@ynao.ac.cn}

\begin{abstract}
We present observational analysis of two successive two-sided loop jets observed by the ground-based New Vacuum Solar Telescope (NVST) and the space-borne {\em Solar Dynamics Observatory} ({\em SDO}). The two successive two-sided loop jets manifested similar evolution process and both were associated with the interaction of two small-scale adjacent filamentary threads, magnetic emerging and cancellation processes at the jet's source region. High temporal and high spatial resolution observations reveal that the two adjacent ends of the two filamentary threads are rooted in opposite magnetic polarities within the source region. The two threads approached to each other, and then an obvious brightening patch is observed at the interaction position. Subsequently, a pair of hot plasma ejections are observed heading to opposite directions along the paths of the two filamentary threads, and with a typical speed of two-sided loop jets of the order \speed{150}. Close to the end of the second jet, we report the formation of a bright hot loop structure at the source region, which suggests the formation of new loops during the interaction. Based on the observational results, we propose that the observed two-sided loop jets are caused by the magnetic reconnection between the two adjacent filamentary threads, largely different from the previous scenario that a two-sided loop jet is generated by magnetic reconnection between an emerging bipole and the overlying horizontal magnetic fields.

\end{abstract}
\keywords{Sun: activity --- Sun: filaments, prominences --- Sun: flares --- Sun: magnetic fields}

\section{Introduction}
Coronal jets are collimated hot plasma flows along open magnetic field lines. They have been discovered in 1980s and later on intensively studied in 1990s thanks to the unprecedented X-ray observations by the {\em Yohkoh} \citep{ogawara91} satellite \citep[e.g.,][]{shibata92,shibata94a,innes97,shimojo96,shimojo98}. Subsequently, the huge and great improvements in the observations taken by many space telescopes such as {\em Hinode} \citep{kosugi07}, {\em Solar TErrestrial RElations Observatory} \citep[{\em STEREO};][]{kaiser08},  {\em Solar Dynamics Observatory} \citep[{\em SDO};][]{pesnell12}, and {\em Interface Region Imaging Spectrograph} \citep[{\em IRIS};][]{depontieu14}, have helped solar physicists achieving important findings and understanding on both observational and theoretical research aspects on coronal jets in recent years \citep[see the recent reviews by e.g.][and references therein]{tsiropoula12,raouafi16} . Since coronal jets or jet-like phenomena such as spicules, surges, and sprays are ubiquitous in the solar atmosphere and always associated with magnetic reconnection process, they are considered to be one of the possible source for the enigmatic problems of coronal heating and acceleration of the fast solar wind \citep[e.g.,][]{innes97,shibata07,tian14,mart17}. Despite the immense advances of the recent few years for enriching our understanding of the different physical, observational processes and numerical investigations behind solar coronal jets, there are still many unsolved problems especially about their driving and evolution mechanisms.

It has been widely accepted that coronal jets are formed due to the reconnection between emerging magnetic fluxes with the ambient magnetic fields \citep[e.g.,][]{shibata94b,liu04,jiang07,yang11,shen14,li15,li17}. Sometimes magnetic flux cancellation in the source region is also important for producing coronal jets \citep[e.g.,][]{chen09,hong11,hong14,hong17,chen15,shen12, adams14,zhang14a,zhang14b,panesar16}. In addition, some authors believe that coronal jets can be launched by the mechanism of chromospheric evaporation \citep[e.g.,][]{shimojo01,miyagoshi03}. \cite{shibata94a} suggested that coronal jets can be divided into two types according to their morphology, namely the so-called ``anenome'' jet and the dubbed ``two-sided loop'' jet. The former appear to frequently occur in or near active regions, while the latter often take place in quiet-Sun regions. The anenome jets are formed due to the reconnection between emerging bipoles and the vertical or oblique ambient magnetic fields. Eventually, the upward and downward hot plasma flows from the reconnection site form the observed jet and the brightening at the jet base, respectively. The two-sided loop jets are thought to be produced by the reconnection between emerging bipoles with the above existing horizontal magnetic field lines, while the hot plasma flows along the horizontal coronal loop away from both sides of the emerging bipoles forms the two-sided loop type jet \citep{yokoyama95,yokoyama96}. It is worth noticing here that observational investigations show the anenome jets to be the most frequent \citep[e.g.,][]{shimojo96,shimojo98,jiang08,chen12,kayshap13,hong13,hong16,lee13,lee15,innes16,zhang16}, whereas the two-sided loop jets type remains very scarcely explored. \cite{yokoyama95} pointed out that the two-sided loop jets are not necessarily observed as jets, they are more likely to be observed as transient loop brightenings without detectable metric type \uppercase\expandafter{\romannumeral3} bursts \citep{kundu98,kundu99}. \cite{alexander99} reported a two-sided loop jet occurred in an active region and they emphasized that the nature of the ambient magnetic field with which the emerging flux interacts affects and hence determines the form of the observed jet. Recently, \cite{jiang13} presented a case of recurrent two-sided-loop jets formed by the reconnection between an emerging bipole with the overlying transequatorial loops. The authors estimated the period and speed of the recurrent jets to be 12 minutes and \speed{128 -- 269}, respectively. \cite{ning16} also reported the observations of two-sided plasma flows in a C-class flare with a quasi-period of about 3 minutes and a speed of \speed{230}. The above cited studies all interpreted the two-sided loop jets as due to the magnetic reconnection between the emerging bipole and the overlying horizontal coronal loops, supporting and reinforcing the existing two-sided loop jet model proposed by \cite{shibata94a} and \cite{yokoyama95}.

In this paper, we present observational analysis of two successive two-sided loop type jets that were observed by the New Vacuum Solar Telescope \citep[NVST;][]{liu12,liu14,xiang16} and {\em SDO} on April 16, 2014. We find that the two-sided loop jets were caused by the reconnection between two adjacent filamentary threads, largely different from the previous most accepted scenario of two-sided loop jets being formed by the reconnection between emerging bipoles and the overlying horizontal coronal loops \citep{shibata94a,yokoyama95}. The used instruments and observational data set are briefly introduced in Section 2. The analysis and the results are described in Section 3. Conclusions and discussions are highlighted in the last section.

\section{Observations}
The NVST images the solar photosphere and chromosphere in TiO-band, G-band, and H$\alpha$ wavelengths. In the present study, we mainly use the H$\alpha$-center data that has a cadence of 12 s and a spatial resolution of 0.3\arcsec. For the H$\alpha$ observation, the observing field-of-view (FOV) of the NVST is about $180\arcsec \times 180\arcsec$ that is large enough to cover a normal active region. Due to the influence of the Earth's turbulent atmosphere, the raw solar images taken by the NVST are reconstructed using high-resolution imaging algorithms \citep{liu14}. We also use the coronal observations taken by the Atmospheric Imaging Assembly \citep[AIA;][]{lemen12} together with magnetic field information from Helioseismic and Magnetic Imager \citep[HMI;][]{schou12} on board the {\em Solar Dynamics Observatory} \citep[{\em SDO};][]{pesnell12}. The AIA has seven extreme ultraviolet (EUV) and three ultraviolet (UV)-visible channels, and the cadence and pixel resolution of the EUV (UV-visible) images are 12 (24) s and 0\arcsec.6, respectively. The cadence of the HMI line-of-sight (LOS) magnetograms is 45 s, and the measurement precision is 10 Gauss. All images used in this paper are differentially rotated to the reference time of 06:30:00 UT.

\section{Results}
\begin{figure}
\epsscale{0.8}
\figurenum{1}
\plotone{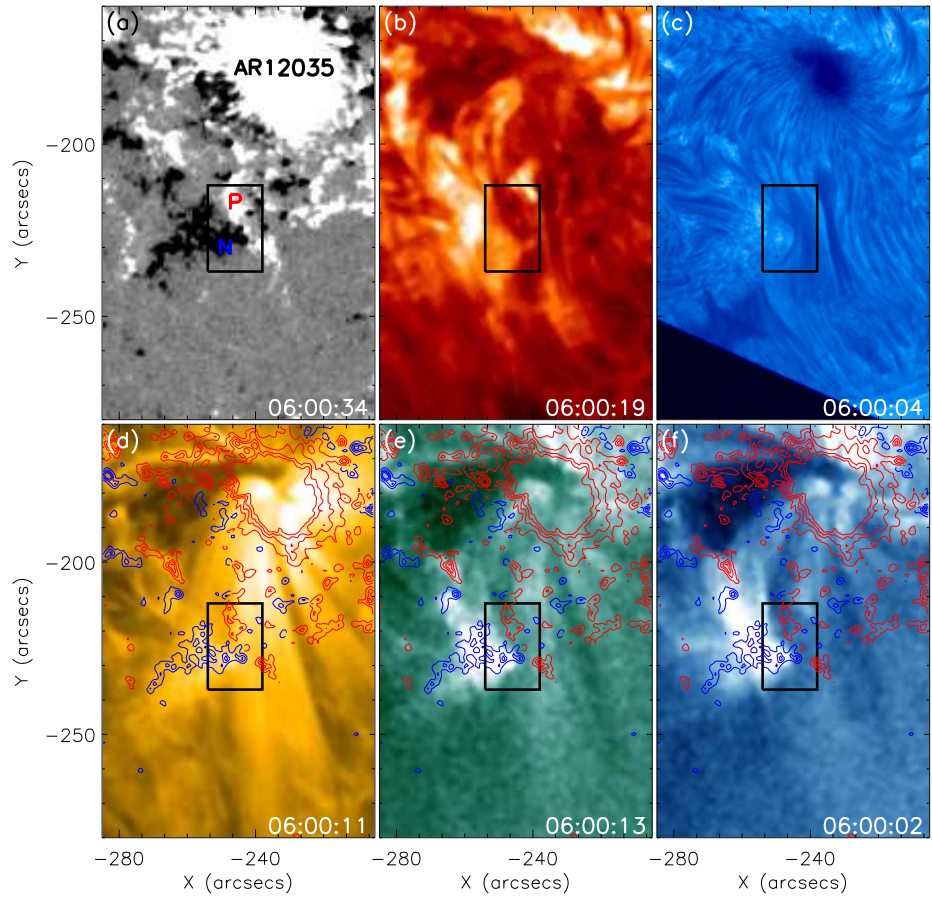}
\caption{An overview of the event before the successive jets. (a) is an HMI LOS magnetogram, in which white (black) represents positive (negative) polarity. (b) and (c) show the chromospheric observations of the source region with AIA 304 \AA and NVST H$\alpha$-center images, respectively. (d) -- (f) are AIA 171, 94, and 335 \AA images overlaid with contours of HMI LOS magnetogram at 06:00:34 UT, and the contour levels are $\pm 50, \pm 100, and \pm 200$, with red (blue) color for positive (negative ) polarity. The black box overlaid in each panel indicate the source region of the successive jets.The FOV of each panel is $80\arcsec \times 120\arcsec$.
\label{fig1}}
\end{figure}

The two successive two-sided loop jets subject of our present study occurred on April 16, 2014 at the southern periphery of NOAA active region 12035. From the {\em GOES} soft X-ray database there is no flare associated with this active region from 05:50:00 UT to 07:10:00 UT. An overview of the observations prior to the jets is shown in \nfig{fig1}. The magnetic source region of the jets is clearly a bipole region simply consisted by a negative and a positive magnetic polarities (the box region in \nfig{fig1} (a)). The same region is also indicated in the other panels. From the NVST H$\alpha$-center and AIA 304 \AA\ images, one can recognize a small filament laid on the magnetic polarity inversion line, while the source region of the jet appears as a bright point in the AIA 94 and 335 \AA\ images. In the AIA 171 \AA image, the source region is obscured by a bunch of coronal loops rooted in active region 12035 (see \nfig{fig1} (d)).

\begin{figure}[!t]
\epsscale{0.8}
\figurenum{2}
\plotone{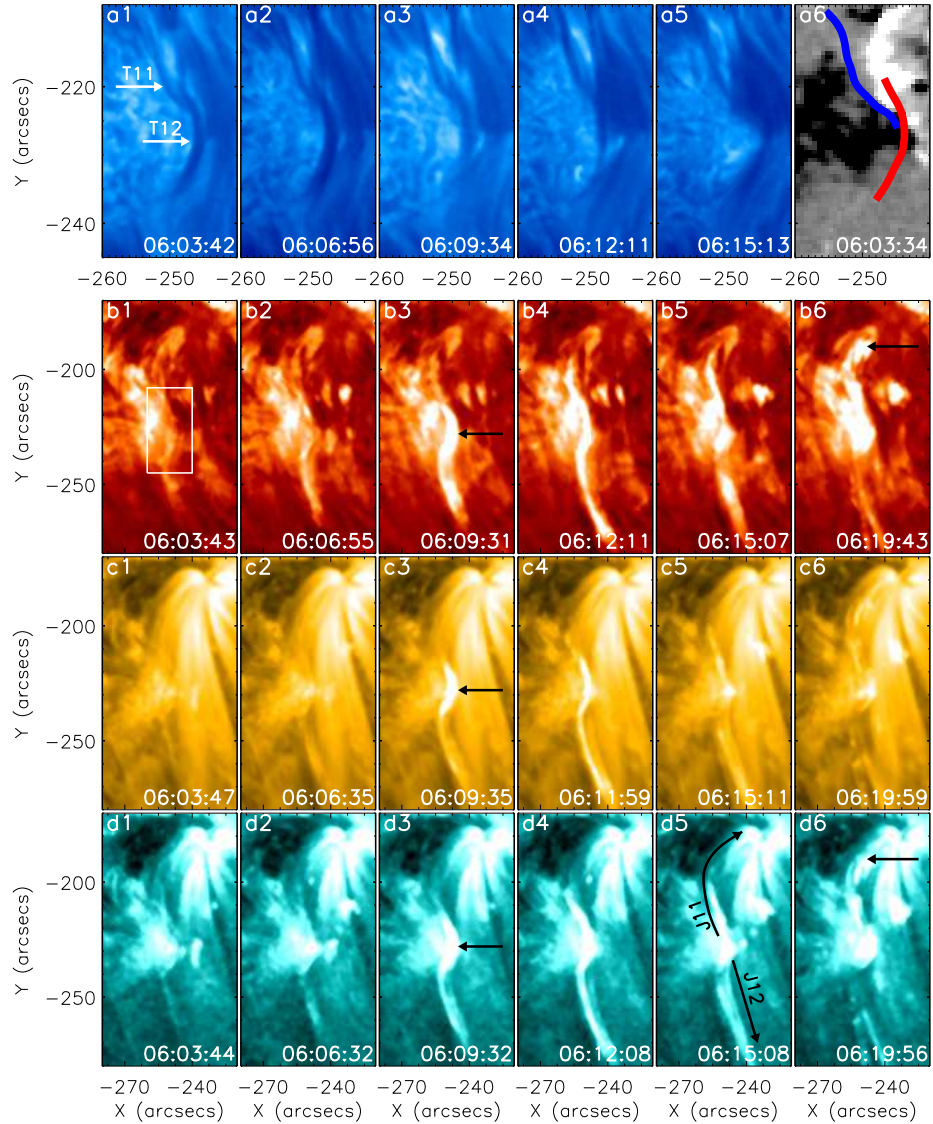}
\caption{The detailed evolution of the first jet. (a1) -- (a5) are NVST H$\alpha$-center images; (a6) is an HMI LOS magnetogram overlaid with the paths of the two filamentary threads determined from (a1); (b1) -- (b6), (c1) -- (c6), and (d1) -- (d6) are AIA 304, 171, and 131 \AA\ time sequence images. T11 and T21 in (a1) label the two small filamentary threads. The two arrows in (d5) indicate the two-sided loop jets.  The black arrows in panels (b3), (c3), and (d3) indicate the brightening at the very beginning of the first jet. The arrows in (b6) and (d6) indicate the rebounding ejection hot plasma from the active region. The FOVs are $20\arcsec \times 37\arcsec$ and $60\arcsec \times 110\arcsec$ for the top row and other rows, respectively. Animations of different wavelength observations are available in the online journal.
\label{fig2}}
\end{figure}

The detailed ejection process of the first jet is displayed in \nfig{fig2} using the NVST H$\alpha$-center and AIA EUV images. To better show the evolution process of the jet and the magnetic field in the source region, the H$\alpha$ images and HMI LOS magnetogram are zoomed in and plotted in the top row panels. It appears evident that before the jet the small filament is in fact composed by two thin adjacent filamentary threads marked as T11 and T12 in \nfig{fig2} (a1). Additionally, the paths of the filamentary threads are highlighted on the HMI LOS magnetogram in \nfig{fig2} (a6). One can see that T11 is clearly recognized to be rooted in the negative magnetic field region (blue), while T12 is rooted in the positive magnetic field region (red). Around 06:06:56 UT, we note that the two threads approached to each other and T12 became thicker and darker than before. A few minutes later, T12 disappeared from the H$\alpha$-center images. In the AIA EUV images, the ejection process of the jet is more obviously. Moreover, around 06:09:31 UT, significant brightening is observed at the cross position of T11 and T12 (marked by arrows in \nfig{fig2} (b3), (c3), and (d3)). This indicates that magnetic reconnection most likely occurred between the two filamentary threads due to their interaction. The absence of the brightening in the H$\alpha$ image probably suggests that the magnetic reconnection site is higher in the corona. In EUV observations, the plasma ejection from the source region is evident along the trajectory of T12, while in the opposite direction a relative weak jet could also be identified along T11. The two-sided loop jet is indicated in \nfig{fig2} (d5) by the two black arrows labeled as J11 and J12. The strong ejection of J12 in EUV images is not observed in the H$\alpha$ images. This is probably due to higher temperature of the jet plasma or simply due to the upward motion of the jet. Because of the lack of H$\alpha$ off-band observations, we can not confirm the reason for the disappearance of T12 from the H$\alpha$-center images. It should be noted that although J11 is very weak, we can however confirm it from the EUV observations by inspecting the bouncing process of the ejecting plasma along T11, which is indicated by the black arrows in \nfig{fig2} (b6) and (d6). Therefore, the whole evolution process confirms that this event is a two-sided loop type jet caused by the reconnection between the two filamentary threads. Here, the formation of the bright mass ejection from the north end of T11 can be explained in terms of bounce back of the northward moving jet (J11) and the evaporation of the heated chromosphere plasma due to the heating of the thermal energy produced by the magnetic reconnection in the jet source region.

\begin{figure}[!t]
\epsscale{0.8}
\figurenum{3}
\plotone{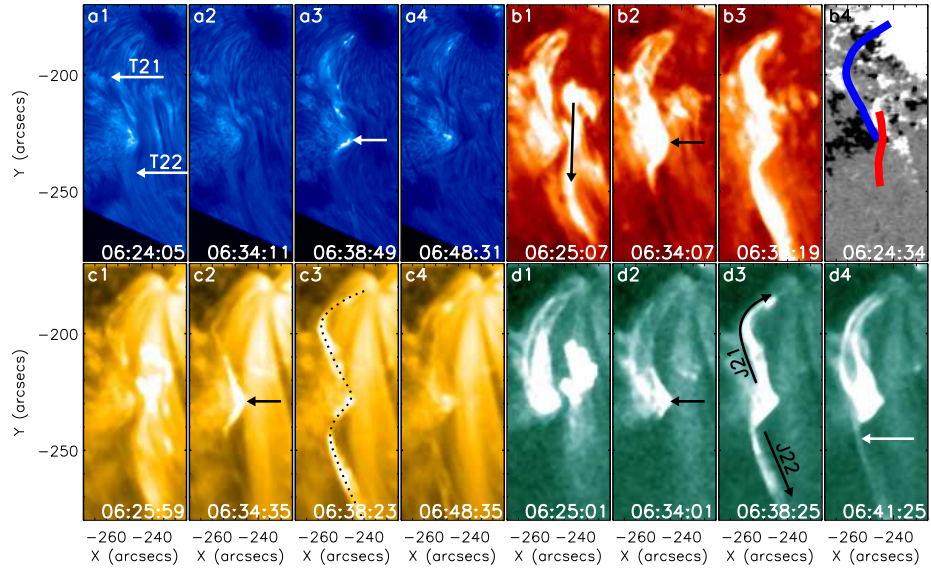}
\caption{The detailed evolution of the second jet. (a1) -- (a4) are NVST  H$\alpha$-center images; (b1) -- (b3), (c1) -- (c4), and (d1) -- (d4) are AIA 304, 171, and 94 \AA\ images, respectively; (b4) is an HMI LOS magnetogram overlaid with the paths of the two filamentary threads determined from (a1). T21 and T22 in panel (a1) label the two filamentary threads. The arrows in panels (a3), (b2), (c2), and (d2) indicate the brightening at very beginning of the second jet. The arrow in panel (b1) indicates the disturbance impacted on T22. The two arrow in panel (d3) indicate the two-sided loop jet, and the arrow in panel (d4) indicates the new formed loop due to the reconnection between the two filamentary threads. The dotted line in panel (c3) shows the path that is used to obtain the time-distance plots showing in \nfig{fig4}. The FOV of each panel is $45\arcsec \times 110\arcsec$.
\label{fig3}}
\end{figure}

The second jet occurred at the same location but about 15 minutes after the first one. The basic evolution process is similar to the first one, but the relevant characteristics are more clear. The detailed evolution phases are shown in \nfig{fig3}. Similar to the first jet, there are two adjacent filamentary threads that can be identified in the H$\alpha$ images before the initiation of the two-sided loop jet (T21 and T22 in \nfig{fig3} (a1)). The spines of the filamentary threads are also overlaid on the HMI LOS magnetogram at 06:24:34 UT (\nfig{fig3} (b4)). Similar to the first jet, it is clear that the two threads are also rooted in opposite magnetic polarities in the jet magnetic source region. Right before the beginning of the second two-sided loop jet, another small plasma ejection is observed on the west of the source region (the black arrow in \nfig{fig3} (b1)), which impacted on T22 and therefore drive T22 to interact with T21. At about 06:34 UT, the crossing position between T21 and T22 became bright in both H$\alpha$ and AIA EUV observations (the arrows in \nfig{fig3} (b2) and (c2)), implying that a magnetic reconnection scenario has occurred between the two threads due to their interaction. After the beginning of the brightening at the source region, a pair of opposite hot plasma ejections are observed along the trajectories of T21 and T22 (\nfig{fig3} (b3) and (d3)). Moreover, a bright loop is observed above the reconnection site as indicated by the white arrow in \nfig{fig3} (d4). Unlike the two-sided loop jet that can be recognized at different AIA wavelengths, this bright loop can only be observed at the hotter 131 and 94 \AA\ channels. There are two explanations to the formation of this bright hot loop system. One is that it is possibly the expected reconnected loop generated in the magnetic reconnection process; the second is that it is possibly caused by the evaporation of the chromosphere cold plasma due to the heating of the thermal energy released by the magnetic reconnection in the jet source region.

\begin{figure}[!t]
\epsscale{0.8}
\figurenum{4}
\plotone{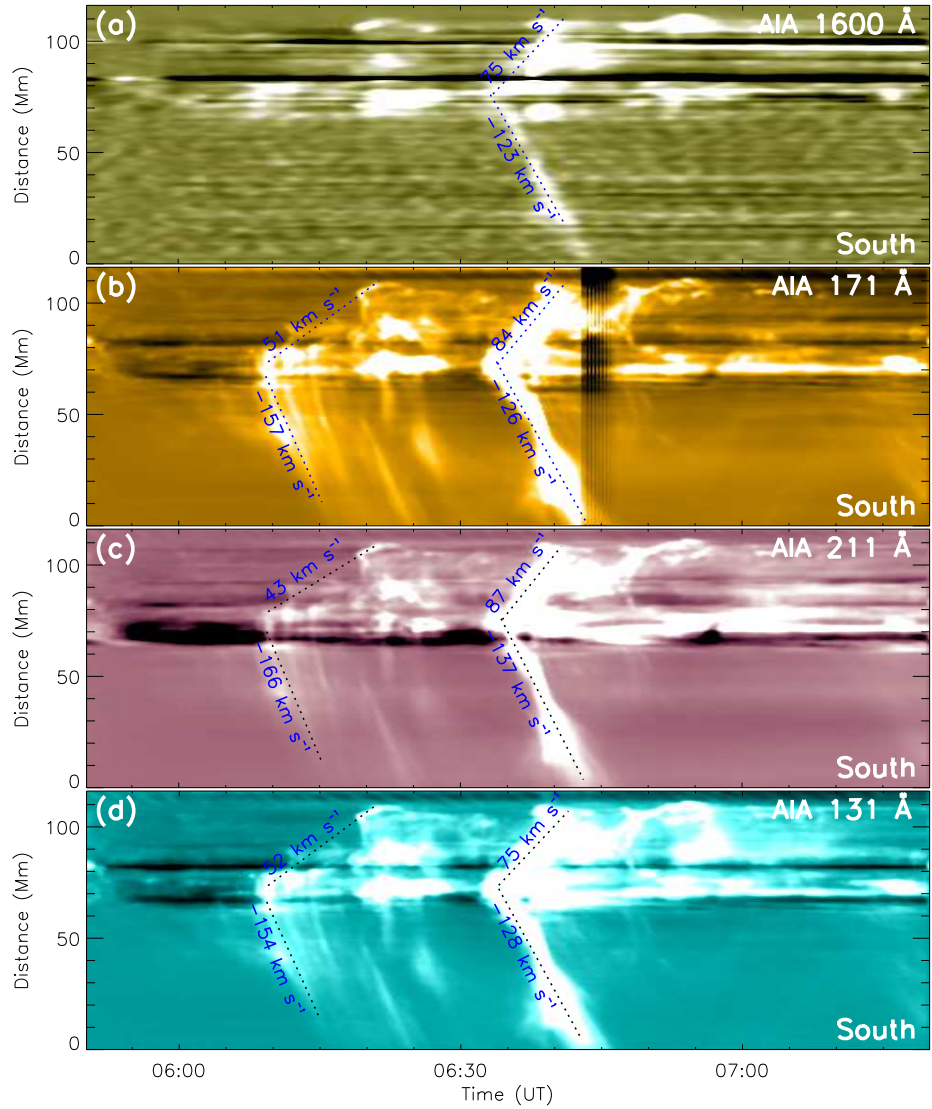}
\caption{Time-distance plots show the ejection of the two successive two-sided loop jets. (a) -- (d) are made from AIA 1600, 171, 211, and 131 \AA\ base-difference images, respectively. The dotted lines are linear fit to the paths of the jets, and the speeds are also plotted in the figure.
\label{fig4}}
\end{figure}

The kinematics of the two successive two-sided loop jets are presented in \nfig{fig4} using time-distance diagrams, which are made along the dotted curve shown in \nfig{fig3} (c3). Since the first jet is relatively weaker than the second one, we can not clearly observe it in the time-distance diagram made from the AIA 1600 \AA\ images. For the second jet, it is clear that the plasma ejections have opposite directions as observed in the direct images. From these time-distance diagrams, we estimate the average ejection speeds of the first two-sided loop jet to be \speed{49 and 159} for J11 and J12, respectively. The average ejection speeds of the second two-sided loop jet are \speed{80 and 128} for J21 and J22, respectively. We find that the ejections towards the southern direction are consistent with previous studies \citep[e.g.,][]{jiang13}, while the ejections toward the northern direction are obviously lower. This is probably because the ejections toward northern direction are trapped in a short closed loop.

\begin{figure}[!t]
\epsscale{0.8}
\figurenum{5}
\plotone{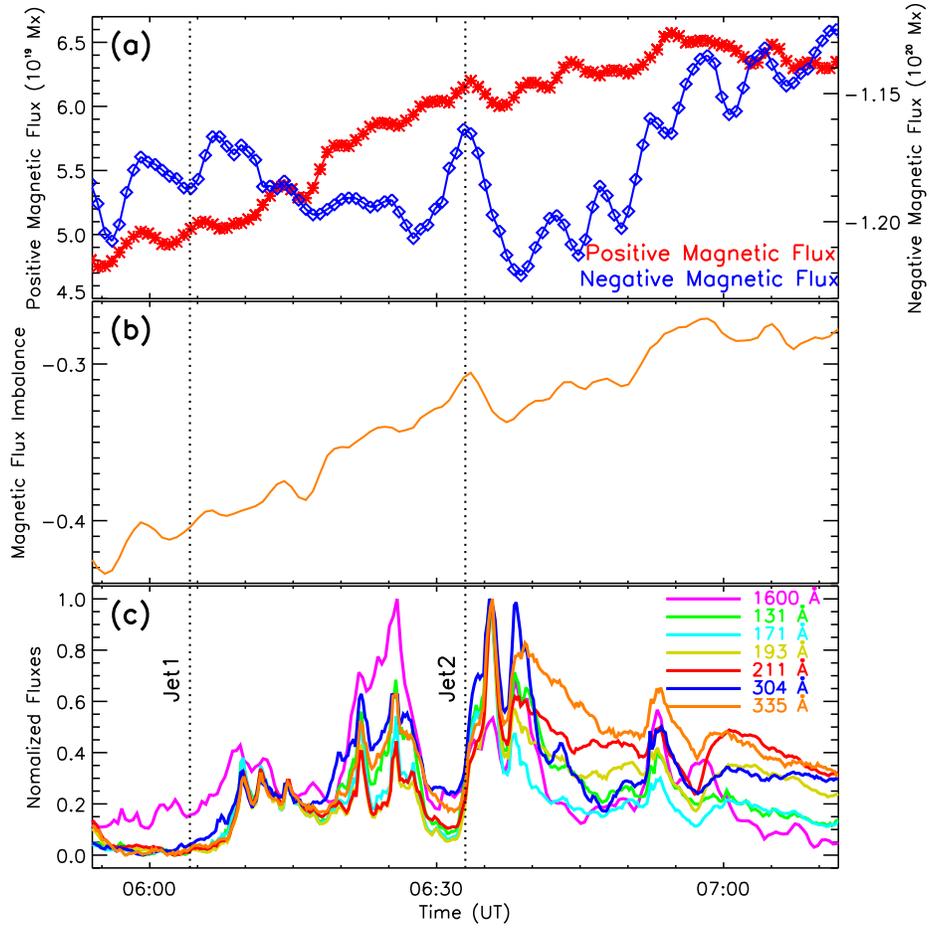}
\caption{The top panel shows the magnetic fluxes in the box region shown in \nfig{fig1} (a). The red and blue lines show the positive and negative magnetic fluxes, respectively. The middle panel shows the magnetic flux imbalance curve defined as $\rm F_{im} = (|F_p| - |F_n|)/(|F_p| + |F_n|)$. Here, $\rm F_{im}$, $\rm F_p$, and $\rm F_n$ are flux imbalance, positive flux and negative flux, respectively. The bottom panel shows the normalized  intensity lightcurves in the same region as the magnetic fluxes. The different channels are shown as different colors. The two vertical dotted lines indicate the start times of the two successive two-sided loop jets, respectively.
\label{fig5}}
\end{figure}

In \nfig{fig5} we report the magnetic fluxes (top panel), magnetic flux imbalance (middle panel), and lightcurves at AIA's different wavelengths (bottom panel), which are obtained from the source region of the jet (the black box in \nfig{fig1} (a)). In the figure, the positive (negative) flux is plotted in red (blue) color, while the lightcurves are plotted in different colors as indicated in the top-right of the bottom panel. The positive magnetic flux displays an increase trend during the time interval of our study, whereas the negative flux shows an irregular variation. When the first jet starts (marked by dotted vertical lines in \nfig{fig5}), both the negative and positive magnetic fluxes are decreased for a while. This may suggest that magnetic cancellation occurred at the jet-base during the initiation stage. For the second jet, the positive magnetic flux decreases a lot, but the negative magnetic flux increases significantly. This may indicate that the negative magnetic flux is emerging, and in the meantime, magnetic cancellation is also occurring between the negative and positive magnetic fluxes. To better present the variation of the net magnetic flux, we show the magnetic flux imbalance in the middle panel in \nfig{fig5}, which is simply defined as the ratio of the net and total magnetic flux. The flux imbalance curve shows decrease and increase trends right after the beginning of the first and second jets, respectively. This again suggests the magnetic cancellation and emerging process around the start of the two jets. The temporal dependence between the jets and the changes of the magnetic fluxes suggest that the observed successive jets are probably related with the detected magnetic flux emerging and cancellations, in agreement with many previous studies \citep[e.g.,][]{shen11,shen12}.

\section{Conclussions \& Discussions}
Using high temporal and high spatial resolution observations taken by NVST and {\em SDO}, we report two successive two-sided loop jets that occurred at the southern periphery of NOAA active region 12035. The results of our analysis suggest that the two successive two-sided loop jets are the consequence of the magnetic reconnection between two adjacent filamentary threads rather than magnetic reconnection between emerging bipoles and their overlying horizontal magnetic fields as proposed in previous model and observations \citep[][]{shibata94a,yokoyama95,jiang13}. Due to magnetic cancellation and emerging within the source region site of the jets, the two adjacent filamentary threads approached to each other and interacted at the crossing area, which triggered the magnetic reconnection between them. The consequence of the magnetic reconnection further resulted in the opposite plasma ejection along the paths of the two filamentary thread and the formation of a hot loop after the second jets. For the first jet, the ejection speeds of the northern and southern plasma ejections are about to be \speed{49 and 159}, respectively. For the second jet, the ejection speeds of the northern and southern plasma ejections are estimated to be \speed{80 and 128}, respectively. The southern ejection speeds of both jets are in agreement with previous observational studies \citep{jiang13}. However, the northern ejection speeds of the two successive jets are relatively lower than the typical speed of two-sided loop jets. This is possibly caused by the fact that the northern ejections are trapped in a closed loop system, where the magnetic field strength and plasma density should be higher than the other ejection path.

\begin{figure}[!t]
\epsscale{0.8}
\figurenum{6}
\plotone{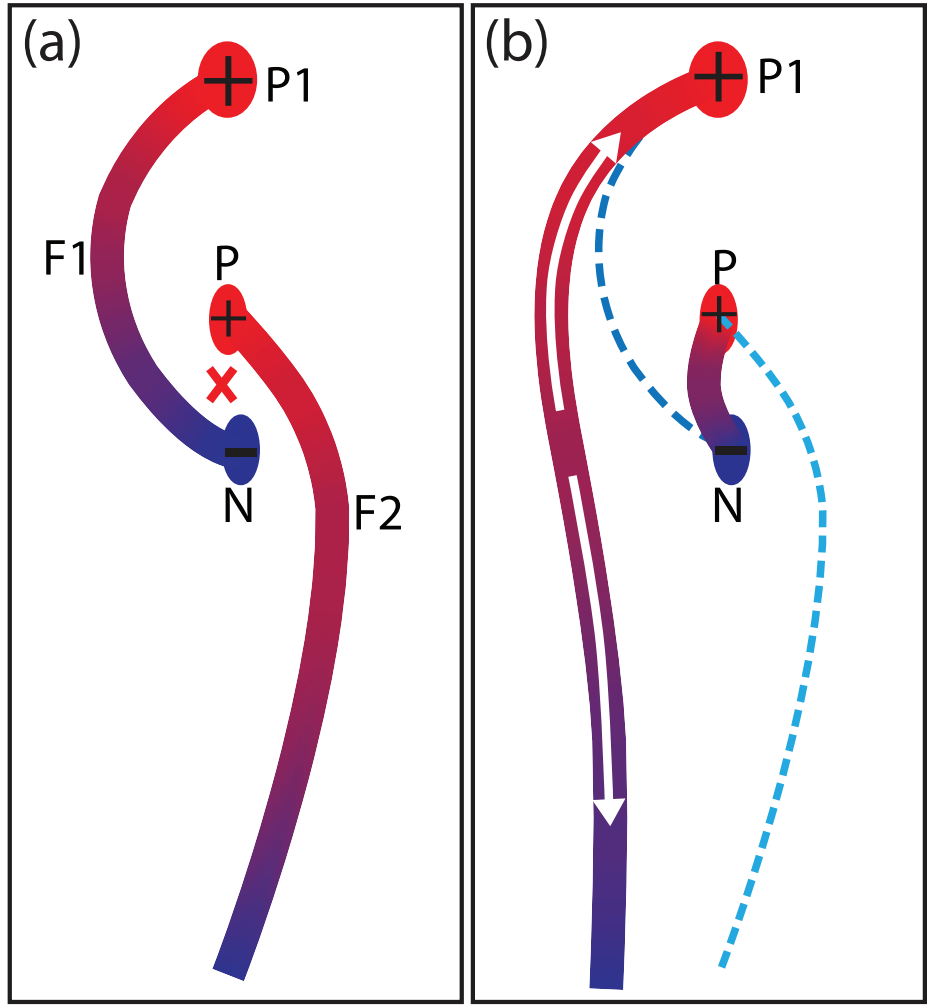}
\caption{A cartoon shows the physical mechanism for producing the two observed two-sided loop jets. Panel (a) shows the magnetic topology before the ejections. The red cross marks the reconnection site between the two filamentary threads. Panel (b) shows the magnetic topology after the magnetic reconnection between the two adjacent filamentary threads, in which the dashed curves indicate the positions of the filamentary threads before the magnetic reconnection. The two white arrows indicate the ejection directions of the two-sided loop jets. Labels P, P1, and N mark the different polarities, in which P and P1 are positive magnetic polarity, N represent negative magnetic polarity.
\label{fig6}}
\end{figure}

The analysis results presented in this paper provide an alternative new triggering mechanism for explaining two-sided loop jets. Based on our analysis results, we present a simple cartoon model in \nfig{fig6} to interpret the observed successive two-sided loop jets. The topological structure before the ejections are shown in \nfig{fig6} (a), in which the left filamentary thread (F1) connects the NOAA active region 12035 (P1) and the negative polarity (N) in the source region of the jet, while the one end of the other filamentary thread (F2) roots in the positive polarity (P) in the source region, and the other end of F2 are assumed to be open or quasi-open. Obviously, such a magnetic topology implies a magnetic null between F1 and F2 (see the red cross symbol in \nfig{fig6} (a)). When F1 and F2 approach to each other due to some disturbances, magnetic reconnection can take place at the crossing position as shown by the red cross symbol. The consequences of the magnetic reconnection are the generation of a pair of opposite plasma ejections along the two filamentary threads, the formation of reconnected magnetic fields, and brightening at the source region. \nfig{fig6} (b) highlights the configuration after the magnetic reconnection, with the blue dashed curves indicating the positions of F1 and F2 before the magnetic reconnection. As can be seen, the reconnected loop joining P and N represents the post-flare-loop observed as a brightening at the source region, while the large one rooting in P1 designates the hot structure observed in the hot AIA 94 \AA\ and 131 \AA\ images after the plasma ejection. The two opposite white arrows indicates the observed two-sided loop jet. It should be pointed out that the plasma ejection from P1 in the first two-sided loop jet and the formation of the hot bright loop are probably caused by the evaporation of the chromosphere cold plasma due to the heating of thermal energy released in the magnetic reconnection process.

To some extent, the jet model presented here share some common characteristics with the so-called tether-cutting model commonly adopted to explain large-scale filament eruptions and coronal mass ejections \citep[][]{moore01,chen14,chen16}. In some studies, the tether-cutting mechanism is also used to explain the formation of flux ropes or filaments \citep[e.g.,][]{cheng15,yan15,yang16,xue17}.  In our present investigation, however, the observed new formed hot loop structure after the two-sided loop jet can possibly be viewed as a new formed flux rope as reported in \cite{cheng15}. These studies including the present one suggest that a common physical mechanism can be used to explain different phenomena on the Sun, and also that small-scale and large-scale solar eruptions events exhibits, to a certain extent, similar observational and physical characteristics. The results of our study are highly encouraging for future investigations, notably in scrutinizing the driving mechanisms of the observed variety of jet-like eruptive solar events.

\acknowledgments We thank the observations provided by NVST and {\em SDO}. This work is supported by the Natural Science Foundation of China (11533009,11403097,11633008), the Yunnan Science Foundation (2015FB191), the Specialized Research Fund for State Key Laboratories, the Open Research Program of the Key Laboratory of Solar Activity of Chinese Academy of Sciences (KLSA201601), the Youth Innovation Promotion Association (2014047) of Chinese Academy of Sciences, and the Key Laboratory of Geospace Environment, CAS, University of Science \& Technology of China. The research by the KSU astronomy unit A. Elmhamdi and A.S. Kordi were supported by King Saud University, Deanship of Scientific Research, College of Science Research Center.

%\begin{comment}

%\end{comment}
%\bibliographystyle{aasjournal}
%\bibliography{two_sided,instrument}

\begin{thebibliography}{}
\bibitem[Adams et al. (2014)]{adams14}
Adams, M., Sterling, A. C., Moore, R. L., \& Gary, G. A. 2014, \apj, 783, 11
\bibitem[Alexander \& Fletcher (1999)]{alexander99}
Alexander, D., \& Fletcher, L. 1999, \solphys, 190, 167
\bibitem[Chen et al. (2009)]{chen09}
Chen, H., Jiang, Y., \& Ma, S. 2009, \solphys, 255, 79
\bibitem[Chen et al. (2014)]{chen14}
Chen, H., Zhang, J., Cheng, X., et al. 2014, \apjl, 797, 15
\bibitem[Chen et al. (2012)]{chen12}
Chen, H.-D., Zhang, J., Ma, S.-L. 2012, RAA, 12, 573
\bibitem[Chen et al. (2015)]{chen15}
Chen, J., Su, J., Yin, Z., et al. 2015, 815, 71
\bibitem[Chen et al. (2016)]{chen16}
Chen, H., Zhang, J., Li, L., \& Ma, S. 2016, \apjl, 818, 27
\bibitem[Cheng et al. (2015)]{cheng15}
Cheng, X., Ding, M. D., \& Fang, C. 2015, \apj, 804, 82
\bibitem[de Pontieu et al. (2014)]{depontieu14}
de Pontieu, B., Title, A. M., Lemen, J. R., et al. 2014, \solphys, 289, 2733
\bibitem[Hong et al. (2011)]{hong11}
Hong, J., Jiang, Y., Zheng, R., et al. 2011, \apjl, 738, 20
\bibitem[Hong et al. (2013)]{hong13}
Hong, J., Jiang, Y., Yang, J., et al. 2013, RAA, 13, 253
\bibitem[Hong et al. (2014)]{hong14}
Hong, J., Jiang, Y., Yang, J., et al. 2014, \apj, 796, 73
\bibitem[Hong et al. (2016)]{hong16}
Hong, J., Jiang, Y., Yang, J., et al. 2016, \apj, 830, 60
\bibitem[Hong et al. (2017)]{hong17}
Hong, J., Jiang, Y., Yang, J., Li, H., \& Xu, Z. 2017, \apj, 835, 35
\bibitem[Innes et al. (2016)]{innes16}
Innes, D. E., Bu\v{c}\'{i}k, Guo, L.-J., \& Nitta, N. 2016, Astron. Nachr., 337, 1024
\bibitem[Innes et al. (1997)]{innes97}
Innes, D. E., Inhester, B., Axford, W. I., \& Wilhelm, K. 1997, \nat, 386, 329
\bibitem[Jiang et al. (2013)]{jiang13}
Jiang, Y., Bi, Y., Yang, J., et al. 2013, \apj, 775, 132
\bibitem[Jiang et al. (2008)]{jiang08}
Jiang, Y., Shen, Y., Bi, Y., Yang, J., \& Wang, J. 2008, \apj, 677, 699
\bibitem[Jiang et al. (2007)]{jiang07}
Jiang, Y. C., Chen, H. D., Li, K. J., Shen, Y. D., \& Yang, L. H. 2007, \aap, 469, 331
\bibitem[Kaiser et al. (2008)]{kaiser08}
Kaiser, M. L., Kucera, T. A., Davila, J. M., et al. 2008, \ssr, 136, 5
\bibitem[Kayshap et al. (2013)]{kayshap13}
Kayshap, P., Srivastava, A. K., \& Murawski, K. 2013, \apj, 763, 24
\bibitem[Kosugi et al. (2007)]{kosugi07}
Kosugi, T., Matsuzaki, K., Sakao, T., et al. 2007, \solphys, 243, 3
\bibitem[Kundu et al. (1999)]{kundu99}
Kundu, M. R., Nindos, A., Raulin, J. P., et al. 1999, \apj, 520, 391
\bibitem[Kundu et al. (1998)]{kundu98}
Kundu, M. R., Raulin, J. -P., Nitta, N., \& Shimojo, M. 1998, \solphys, 178, 173
\bibitem[Li et al. (2017)]{li17}
Li, H., Jiang, Y., Yang, J., et al. 2017, \apj, 836, 235
\bibitem[Lee et al. (2015)]{lee15}
Lee, E. J., Archontis, V., \& Hood, A. W. 2015, \apj, 799, 79
\bibitem[Lee et al. (2013)]{lee13}
Lee, K. -S., Innes, D. E., Moon, Y. -J., et al. 2013, \apj, 766, 1
\bibitem[Lemen et al. (2012)]{lemen12}
Lemen, J. R., Title, A. M., Akin, D. J., et al. 2012, \solphys, 275, 17
\bibitem[Li et al.(2015)]{li15} Li, H.~D., Jiang, Y.~C., Yang, J.~Y., Bi, Y., \& Liang, H.~F.\ 2015, \apss, 359, 4
\bibitem[Liu \& Kurokawa (2004)]{liu04}
Liu, Y., \& Kurokawa, H. 2004, \apj, 610, 1136
\bibitem[Liu et al.(2012)]{liu12} Liu, Y., Xu, Z., \& YNST Team 2012, Hinode-3: The 3rd Hinode Science Meeting, 454, 463
\bibitem[Liu et al. (2014)]{liu14}
Liu, Z., Xu, J., Gu, B.-Z., et al. 2014, RAA, 14, 705
\bibitem[Mart\'{i}nez-Sykora, J. et al. (2017)]{mart17}
Mart\'{i}nez-Sykora, J., De Pontieu, B., Hansteen, V. H., et al. 2017, Science, 356, 1269
\bibitem[Miyagoshi \& Yokoyama (2003)]{miyagoshi03}
Miyagoshi, T., \& Yokoyama, T. 2003, \apjl, 593, L133
\bibitem[Moore et al. (2001)]{moore01}
Moore, R. L., Sterling, A. C., Hudson, H. S., \& Lemen, J. R. 2001, \apj, 552, 833
\bibitem[Ning (2016)]{ning16}
Ning, Z. 2016, \apss, 361, 22
\bibitem[Ogawara et al. (1991)]{ogawara91}
Ogawara, Y., Takano, T., Kato, T., et al. 1991, \solphys, 136, 1
\bibitem[Panesar et al.(2016)]{panesar16} Panesar, N.~K., Sterling, A.~C., Moore, R.~L., \& Chakrapani, P.\ 2016, \apjl, 832, L7
\bibitem[Pesnell et al. (2012)]{pesnell12}
Pesnell, W. D., Thompson, B. J., \& Chamberlin, P. C. 2012, \solphys, 275, 3
\bibitem[Raouafi et al. (2016)]{raouafi16}
Raouafi, N. E., Patsourakos, S., Pariat, E., et al. 2016, \ssr, 201, 1
\bibitem[Schou et al. (2012)]{schou12}
Schou, J., Borrero, J. M., Norton, A. A., et al. 2012, \solphys, 275, 229
\bibitem[Shen et al. (2011)]{shen11}
Shen, Y., Liu, Y., Su, J., \& Ibrahim, A. 2011, \apjl, 735, 43
\bibitem[Shen et al. (2012)]{shen12}
Shen, Y., Liu, Y., Su, J., \& Deng, Y. 2012, \apj, 745, 164
\bibitem[Shen et al. (2014)]{shen14}
Shen, Y., Ichimoto, K., Ishii, T. T., et al. 2014, \apj, 2014, 786, 151
\bibitem[Shibata et al. (2007)]{shibata07}
Shibata, K., Nakamura, T., Matsumoto, T., et al. 2007, Science, 318, 1591
\bibitem[Shibata et al. (1994a)]{shibata94a}
Shibata, K., Nitta, N., Matsumoto, R., et al. 1994a, in X-ray solar physics from Yohkoh, ed. Y. Uchida et al., 29
\bibitem[Shibata et al. (1994b)]{shibata94b}
Shibata, K., Nitta, N., Strong, K. T., et al. 1994b, \apjl, 431, 51
\bibitem[Shibata et al. (1992)]{shibata92}
Shibata, K., Ishido, Y., Acton, L. W., et al. 1992, \pasj, 44, 173
\bibitem[Shimojo et al. (1996)]{shimojo96}
Shimojo, M., Hashimoto, S., Shibata, K., et al. 1996, \pasj, 48, 123
\bibitem[Shimojo et al. (1998)]{shimojo98}
Shimojo, M., Shibata, K., Harvey, K. L., et al. 1998, \solphys, 178, 379
\bibitem[Shimojo et al. (2001)]{shimojo01}
Shimojo, M., Shibata, K., Yokoyama, T., \& Hori, K. 2001, \apj, 550, 1051
\bibitem[Tian et al. (2014)]{tian14}
Tian, H., DeLuca, E. E., Cranmer, S. R., et al. 2014, Science, 346, 1255711
\bibitem[Tsiropoula et al. (2012)]{tsiropoula12}
Tsiropoula, G., Tziotziou, K., Kontogiannis, I., et al. 2012, \ssr, 169, 181
\bibitem[Xiang et al. (2016)]{xiang16}
Xiang, Y.-Y., Liu, Z., \& Jin, Z. -Y. 2016, \na, 830, 60
\bibitem[Xue et al. (2017)]{xue17}
Xue, Z., Yan, X., Yang, L., Wang, J., \& Zhao, L. 2017, \apjl, 840, 23
\bibitem[Yan et al. (2015)]{yan15}
Yan, X. L., Xue, Z. K., Pan, G. M., et al. 2015, \apjs, 219, 17
\bibitem[Yang et al. (2016)]{yang16}
Yang, B., Jiang, Y., Yang, J., Yu, S., \& Xu, Z. 2016, \apj, 816, 41
\bibitem[Yang et al. (2011)]{yang11}
Yang, L.-H., Jiang, Y.-C., Yang, J.-Y., et al. 2011, RAA, 11, 1229
\bibitem[Yokoyama \& Shibata (1995)]{yokoyama95}
Yokoyama, T., \& Shibata, K. 1995, \nat, 375, 42
\bibitem[Yokoyama \& Shibata (1996)]{yokoyama96}
Yokoyama, T., \& Shibata, K. 1996, \pasj, 48, 353
\bibitem[Zhang \& Ji (2014a)]{zhang14a}
Zhang, Q. M., \& Ji, H. S. 2014a, \aap, 561, 134
\bibitem[Zhang \& Ji (2014b)]{zhang14b}
Zhang, Q. M., \& Ji, H. S. 2014b, \aap, 567, 11
\bibitem[Zhang et al. (2016)]{zhang16}
Zhang, Q. M., Li, D., Ning, Z. J., et al. 2016, \solphys, 291, 859
\end{thebibliography}

\end{document}